\documentclass{IEEEcsmag}
\usepackage{balance}
\usepackage{caption}

\usepackage[colorlinks,urlcolor=blue,linkcolor=blue,citecolor=blue]{hyperref}
\usepackage[hyphenbreaks]{breakurl}
\usepackage{upmath,color}
\usepackage{textgreek}

\usepackage{xcolor}
\usepackage{xspace}
\usepackage{subcaption}
\usepackage{pifont}
\usepackage{graphbox}
\usepackage{graphicx}
\usepackage{epstopdf}
\usepackage{dblfloatfix} % To enable figures at the bottom of page
\usepackage[htt]{hyphenat}
\let\rfoot\relax
\usepackage{fancyhdr}
\usepackage{setspace}
\usepackage{relsize}
\usepackage{textcomp}
\usepackage{circledsteps}
\usepackage{url}
\usepackage[htt]{hyphenat}
\usepackage[super]{nth}
\usepackage{comment}
\usepackage{transparent}
\usepackage{nicematrix,booktabs}

\usepackage[T1]{fontenc}
\usepackage{lipsum}
\usepackage[normalem]{ulem}
\makeatletter
\def\SOUL@hlpreamble{%
\setul{\dimexpr\dp\strutbox-2pt}{\dimexpr\ht\strutbox+\dp\strutbox-2pt\relax}
\let\SOUL@stcolor\SOUL@hlcolor
\SOUL@stpreamble
}
\makeatother

\newcommand\khlc[1][yellow]{
  \bgroup
  \markoverwith{\textcolor{#1}{\rule[-.5ex]{1pt}{2.5ex}}}
  \ULon
}

\newcommand{\quotes}[1]{``#1''}

\newcommand{\todo}[1]{\bgroup\color{white}\textbf{\khlc[black]{TODO: [#1]}}\egroup\xspace}
\newcommand{\fixme}[1]{\bgroup\color{red}\textbf{\khlc{FIXME: [#1]}}\egroup\xspace}
\newcommand{\pointer}[1]{\bgroup\color{white}\textbf{\khlc[red]{POINTER: [#1 is working here]}}\egroup\xspace}
\newcommand{\red}[1]{\bgroup\color{red}#1\egroup\xspace}
\newcommand{\blue}[1]{\bgroup\color{blue}#1\egroup\xspace}
\newcommand{\reviewer}[1]{\bgroup\color{blue}#1\egroup\xspace}
\newcommand{\ranswer}[1]{\bgroup\color{red}#1\egroup\xspace}
\let\labelindent\relax
\usepackage{enumitem}
\usepackage{amssymb}
\usepackage[framemethod=tikz]{mdframed}
\usepackage[numbers,sort&compress]{natbib}
\tikzset{
    vertical align/.style={
        baseline=-.5*(height("$+$")-depth("$+$"))
    }
}

\newcommand*{\circleNum}[1]{\tikz[baseline=(char.base)]{
              \node[shape=circle,fill,inner sep=0.9pt] (char) {\textcolor{white}{\scriptsize{#1}}};}}
\usepackage{calc}
\usepackage{amsmath,algorithm,algpseudocode}
\usepackage{cases}
\algrenewcommand\algorithmicindent{0.5em}
\makeatletter
\algrenewcommand\ALG@beginalgorithmic{\footnotesize}
\makeatother
\makeatletter
\renewcommand{\Function}[2]{%
  \csname ALG@cmd@\ALG@L @Function\endcsname{#1}{#2}%
  \def\jayden@currentfunction{#1}%
}
\newcommand{\funclabel}[1]{%
  \@bsphack
  \protected@write\@auxout{}{%
    \string\newlabel{#1}{{\jayden@currentfunction}{\thepage}}%
  }%
  \@esphack
}
\makeatother

\usepackage{pythonhighlight}
\definecolor{codegray}{rgb}{0.5,0.5,0.5}
\lstdefinestyle{pythonStyle}{
  basicstyle=\tiny\ttfamily\footnotesize\linespread{0.5},
  basewidth = {.54em},
  commentstyle=\color{codegray},
  frame=single,
  language=Python,
  stepnumber=1,
  numbers=left,
  numbersep=5pt,
  numberstyle=\tiny\color{codegray},
  tabsize=1,
  showspaces=false,
  showstringspaces=false,
  breaklines=false,
  mathescape,
  keywordstyle={\color{black}},
  emph={Load, GraphPre, BatchPre, Aggr, Trans},
  emphstyle={\bfseries\color{orange}},
  moredelim=**[is][\color{red}]{~}{~},
  moredelim=**[is][\color{blue}]{<}{>},
  moredelim=**[is][\color{orange}]{@}{@},
  literate={\\~}{{\textasciitilde}}1
  {\\<}{{\unichar{"003C}}}1
  {\\>}{{\unichar{"003E}}}1
  {\\@}{{\unichar{"0040}}}1
}

\usepackage{tabularx}
\usepackage{array}
\usepackage{multirow}
\usepackage{booktabs}
\usepackage[para]{threeparttable}
\usepackage{colortbl}
\usepackage{multirow}

\makeatletter
\def\hlinewd#1{%
\noalign{\ifnum0=`}\fi\hrule \@height #1 %
\futurelet\reserved@a\@xhline}
\makeatother
\usepackage{hhline}
\newcolumntype{C}{>{\centering\arraybackslash}X}
\usepackage[framemethod=tikz]{mdframed}
\usepackage{marginnote}
\usepackage{pdfcomment}
\mdfsetup{
  hidealllines=true,
  innerleftmargin=2pt,
  innerrightmargin=2pt,
  innertopmargin=2pt,
  innerbottommargin=2pt,
  leftmargin=-2pt,
  rightmargin=-2pt,
  skipabove=-2pt,
  skipbelow=-2pt,
  backgroundcolor=blue!20}

\newlength{\markerHeight}
\newlength{\markerMargin}
\newlength{\linespace}
\newlength{\linedepth}

\sbox0{yf}
\definecolor{mylime}{RGB}{205, 220, 57}
\definecolor{mygreen}{RGB}{60, 200, 0}
\definecolor{myblue}{RGB}{0, 51, 204}

\colorlet{soulred}{red!20}
\colorlet{soulgreen}{green!20}
\colorlet{soulblue}{blue!20}

\usepackage{enumitem}
\setlistdepth{5}
\renewlist{itemize}{itemize}{5}
\setlist[itemize,1]{label=$\bullet$}
\setlist[itemize,2]{label=$\circ$}
\setlist[itemize,3]{label=$\ast$}
\setlist[itemize,4]{label=-}
\setlist[itemize,5]{label=$\cdot$}

\makeatletter
\def\SOUL@hlpreamble{%
\setul{\dimexpr\dp\strutbox-2pt}{\dimexpr\ht\strutbox+\dp\strutbox-2pt\relax}
\let\SOUL@stcolor\SOUL@hlcolor
\SOUL@stpreamble
}
\makeatother

\jtitle{This manuscript is an extended version of the original paper accepted by IEEE Micro.}
\pubyear{2025}
\begin{document}

% \sptitle{THEME ARTICLE: CACHE COHERENT INTERCONNECTS AND RESOURCE DISAGGREGATION TECHNIQUES}

\title{Containerized In-Storage Processing and Computing-Enabled SSD Disaggregation}

\author{Miryeong Kwon$^{*}$, Donghyun Gouk$^{*}$, Eunjee Na$^{*}$, Jiseon Kim$^{*}$, Junhee Kim$^{*}$, Hyein Woo$^{*}$, Eojin Ryu$^{*}$, Hyunkyu Choi$^{*}$, Jinwoo Baek$^{*}$, Hanyeoreum Bae$^{*}$, Mahmut Kandemir$^{\ddagger}$, Myoungsoo Jung$^{*\psi\dagger}$}
\affil
{
\\
\\$^{*}$Next-Generation Silicon and Research Division, \textbf{Panmnesia, Inc.}, Daejeon, South Korea
\\$^{\psi}$Advanced Product Engineering Division, \textbf{Panmnesia, Inc.}, Seoul, South Korea
\\$^\ddagger$Pennsylvania State University, Pennsylvania, United States
\\$^\dagger$KAIST, Daejeon, South Korea
}

% \footernote{This manuscript is an extended version of our paper presented at IEEE Micro.}
% \markboth{THEME ARTICLE: CACHE COHERENT INTERCONNECTS AND RESOURCE DISAGGREGATION TECHNIQUES}{THEME ARTICLE: CACHE COHERENT INTERCONNECTS AND RESOURCE DISAGGREGATION TECHNIQUES}

%-------------------------------------------------------------------------------
\begin{abstract}
ISP minimizes data transfer for analytics but faces challenges in adaptation and disaggregation. We propose \emph{DockerSSD}, an ISP model leveraging OS-level virtualization and lightweight firmware to enable containerized data processing directly on SSDs. Key features include \emph{Ethernet over NVMe} for network-based ISP management and \emph{Virtual Firmware} for secure, efficient container execution. DockerSSD supports disaggregated storage pools, reducing host overhead and enhancing large-scale services like LLM inference. It achieves up to 2.0$\times$ better performance for I/O-intensive workloads, and 7.9$\times$ improvement in distributed LLM inference.
\end{abstract}
%-------------------------------------------------------------------------------

\maketitle
%-------------------------------------------------------------------------------
\label{sec:introduction}
\chapteri{I}n-storage processing (ISP) is an emerging storage model widely adopted for data analytics, enabling efficient exploration of large datasets by minimizing data transfer overhead between the host and storage. Recent industry proposals advocate disaggregating ISP resources into storage arrays or nodes, significantly reducing server-side computational demands. To realize this vision, researchers explored integrating data processing capabilities directly into SSDs \cite{kang2013enabling, lee2014accelerating, seshadri2014willow, gu2016biscuit, kim2016storage, jo2016yoursql, jin2017kaml, koo2017summarizer, jun2018grafboost, torabzadehkashi2019catalina, ruan2019insider, mansouri2022genstore, gouk2023containerized, gouk2024dockerssd}. However, adapting SSDs to meet various application requirements remains a significant challenge.

The primary challenge in designing a flexible ISP model and storage disaggregation lies in creating practical runtimes and APIs, rather than in the computational capabilities of modern SSDs. Protecting vendor-sensitive data and intellectual property restricts the disclosure of internal hardware and firmware, making it essential for SSD vendors to provide ISP runtimes and APIs. However, evolving application requirements complicate the design process, as substantial source-level modifications are needed to offload existing applications. New host-side daemons must also be developed to interact with and synchronize the disaggregated storage, further increasing complexity.

Introducing intelligence at the storage level introduces challenges in managing I/O requests in block-device format while simultaneously processing data near the flash. This dual functionality can create vulnerabilities and resource protection issues, making SSDs unreliable for ISP tasks and user operations. For instance, SSDs lack awareness of file management and block layout, allowing host-side users to modify in-storage data during processing, potentially leading to unpredictable results. Given these limitations, ISP is currently confined to specific data-processing applications, such as key-value or object SSDs \cite{kvssd, jin2017kaml, im2020pink, issd, kwon2022vigil, duffy2023dotori, park2023kv}, and has not been partially applied to resource disaggregation.

We present \emph{DockerSSD}, an adaptable ISP model capable of diverse data processing near the flash without requiring source-level modifications. DockerSSD implements OS-level virtualization within SSDs, enabling ISP to operate in a containerized manner. This design integrates storage intelligence into existing computing environments, accelerating decision-making. Users can run algorithms without modifying them for vendor-specific runtimes or using a specialized toolchain. This enables individual computational SSDs to function independently, supporting their disaggregation from the host and formation into a computing-enabled storage pool. 

While OS-level virtualization promotes ISP adoption, two challenges arise when integrating ISP with containers. First, containers, being self-contained and independent, require a specialized interface for managing flash-based data processing, compatible with the storage and network stacks. Second, their encapsulated nature demands a firmware redesign to dynamically construct containers and provide an execution environment. To address these, we propose a communication framework for ISP that uses Ethernet and firmware to autonomously download and execute Docker images. We also redesign firmware to virtualize the ISP environment, enabling data processing without application-level changes. This approach simplifies host management of disaggregated storage and supports large-scale pre-trained model services like large language models (LLMs).

\noindent \textbf{A novel communication method for ISP.}
We propose \emph{Ethernet over NVMe} (Ether-oN), a kernel driver enabling network-based ISP management and direct communication between host users and SSDs. Ether-oN introduces asynchronous upcalls and packet overriding within the NVMe protocol to support Ethernet networking. This allows our ISP model to leverage command-line interfaces from Docker. With Ether-oN, users can supply data, monitor ISP statuses, and retrieve results directly from the SSD, enabling real-time, on-demand data analysis. 
Since Ether-oN requires no modifications to existing interfaces, it parallelizes the network and block I/O operations.

\noindent \textbf{Containerizing ISP with firmware.}
We present \emph{Virtual Firmware} (Virtual-FW), a lightweight firmware stack that integrates minimal OS functionality and a container environment into the SSD. By emulating system calls on bare-metal hardware, Virtual-FW maintains ISP system call execution costs comparable to function management costs. 
It creates \emph{ISP-containers} from existing Docker images and executes them without the overhead of a full OS. Virtual-FW ensures secure and portable processing by employing NVMe namespace and memory isolation to safeguard SSD resources.

\noindent \textbf{Disaggregating computing-enabled storage.}
Using Ether-oN and Virtual-FW, DockerSSD is equipped with its network IP, enabling uninterrupted computation and the creation of a computing-enabled storage pool. 
This disaggregated pool significantly reduces host-side computation overhead and provides distributed data processing capabilities, improving a wide range of large-scale services. 
We demonstrate its effectiveness through a case study on distributed inference for diverse LLM models \cite{azizi2024lamda, floridi2020gpt, lieber2021jurassic, ren2023pangu, rae2021scaling, sejnowski2023large, chowdhery2023palm, narayanan2021efficient}. 
The storage pool achieves substantial performance gain by eliminating data movement and minimizing memory requirements during processing.

We validate DockerSSD using a PCIe SSD prototype with an NVMe controller \cite{jung2020openexpress} and a multi-core processor on a 16nm FinFET FPGA. 
DockerSSD outperforms host system and leading ISP models \cite{seshadri2014willow, gu2016biscuit, shadley2018deployment} by 1.3$\times$ and 1.8$\times$, respectively. 
In addition, the compute-enabled storage pool enhances LLM performance by an average of 7.9$\times$.

\vspace{-8pt}
%-------------------------------------------------------------------------------

%-------------------------------------------------------------------------------
\section{PRELIMINARY}
\label{sec:background}
\begin{figure}
    \centering
    % \vspace{0pt}
    \includegraphics[width=1\linewidth]{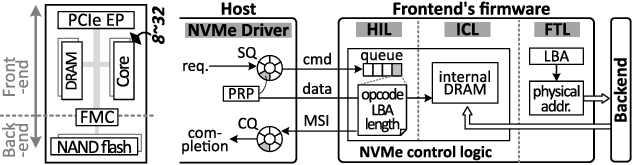}
    \begin{subfigure}{\linewidth}
      \centering
      \begin{tabularx}{\textwidth}{
        p{\dimexpr.50\linewidth-2\tabcolsep-1.3333\arrayrulewidth}% column 1
        p{\dimexpr.50\linewidth-2\tabcolsep-1.3333\arrayrulewidth}% column 2
        }
        \vspace{-5pt} \caption{Storage card.} \label{fig:bck_arch} &
        \vspace{-5pt} \caption{Firmware stack.} \label{fig:bck_nvme}
      \end{tabularx}
    \end{subfigure}
    \vspace{-1pt}
    \caption{Architecture of high-performance SSDs.} \label{fig:bck_ssd}
    %\vspace{5pt}
  \end{figure}
% \begin{figure*}[t]
%   \centering
%   \vspace{-10pt}
%   \includegraphics[width=0.98\linewidth]{figs/dockerssd_view}
%   \begin{subfigure}{0.1\linewidth}
%     \centering \vspace{-50pt}
%     \caption{Stack.} \label{fig:bck_docker}
%   \end{subfigure}
%   \begin{subfigure}{0.38\linewidth}
%     \centering\vspace{-50pt}
%     \caption{Container creation from an image.} \label{fig:bck_image}
%   \end{subfigure}
%   \begin{subfigure}{0.33\linewidth}
%     \centering\vspace{-50pt}
%     \caption{ISP containerization.} \label{fig:backmedia1}
%   \end{subfigure}
%   \begin{subfigure}{0.15\linewidth}
%     \centering
%     \vspace{-50pt}
%     \caption{Media mgmt.} \label{fig:backmedia2}
%   \end{subfigure}
%   \vspace{20pt}
%   \caption{ OS-level virtualization stack and High-level view of DockerSSD.} \label{fig:eval_batch}
% \end{figure*}

\subsection{High Performance SSDs}
\noindent \textbf{Hardware architecture.}
As shown in Figure \ref{fig:bck_arch}, modern SSDs are typically divided into two main components: the \emph{frontend} computing complex and the \emph{backend} storage \cite{jung2013revisiting, matam2019graphssd, jun2018grafboost, kim2023decoupled, sun2023leaftl}. The frontend includes embedded multi-core processors and internal DRAM, connected to a PCIe endpoint and the backend via a high-speed interconnect. The backend consists of multiple I/O buses, or channels, each linking flash packages through a flash memory controller (FMC). In recent years, the frontend has seen significant hardware advancements, enabling SSDs to handle diverse in-storage data processing tasks. Embedded processors now rival low- to mid-range multi-core processors in computational capability, while PCIe throughput approaches host-side system bus bandwidth. For example, modern embedded processors achieve high frequencies (e.g., 2GHz with 8 cores) \cite{issd, layerscapeprocessors} or scale to more cores (500MHz with 32 cores) \cite {microsemipm8609}, providing substantial computational power to the frontend.

\noindent \textbf{NVMe.}
SSDs communicate with a host via NVMe \cite{nvmexpress}, which defines queues, I/O commands, and data transfer methods over PCIe. Each queue comprises a \emph{submission queue} (SQ) and a \emph{completion queue} (CQ). When a user initiates a request, the NVMe driver places it in an SQ and signals the SSD by updating its doorbell register. The NVMe control logic retrieves the command from the SQ, which includes memory pointers called \emph{physical region pages} (PRPs) that indicate the data's location in the host's kernel space. Using PRPs, the control logic transfers the data and processes the request. Upon completion, it updates the CQ entry and notifies the host via a \emph{message signaled interrupt} (MSI), which underpins the handling of host ``block'' semantics.

\noindent \textbf{Firmware stack.}
To bridge the gap between the host's block-based I/O and the flash's page-based storage, SSD firmware in the frontend employs three key layers: the \emph{host interface layer} (HIL), \emph{internal cache layer} (ICL), and \emph{flash translation layer} (FTL). The HIL implements NVMe control logic, analyzing incoming requests to extract key I/O details. The ICL relocates data to internal DRAM, functioning as a memory cache. The FTL maps \emph{logical block addresses} (LBAs) to physical flash addresses, dispatches buffered requests, and ensures backend reliability. These layers are essential for block I/O operations and require precise coordination to support an ISP model.

  \begin{figure}
    \centering
    % \vspace{-2pt}
    \includegraphics[width=1\linewidth]{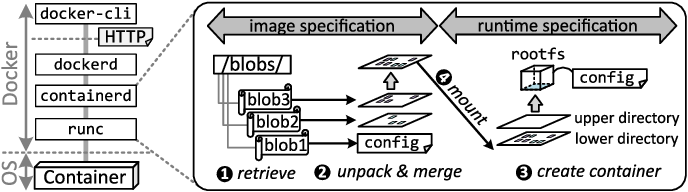}
    \begin{subfigure}{\linewidth}
      \centering
      \begin{tabularx}{\textwidth}{@{}
        p{\dimexpr.35\linewidth-2\tabcolsep-1.3333\arrayrulewidth}% column 1
        @{\hspace*{5pt}}p{\dimexpr.71\linewidth-2\tabcolsep-1.3333\arrayrulewidth}% column 2
        @{}}
        \vspace{-2pt} 
        \caption{Docker stack.} \label{fig:bck_docker} &
        \vspace{-2pt} 
        \caption{Container creation from an image.} \label{fig:bck_image}
      \end{tabularx}
    \end{subfigure}
    \vspace{1pt}
    \caption{OS-level virtualization stack.} \label{fig:bck_virt}
    \vspace{1pt}
  \end{figure}

\vspace{-9pt}
\subsection{OS-Level Virtualization} \label{subsec:docker}

\noindent \textbf{Container and Docker.}
Containers use cgroups \cite{linuxcgroups} and namespaces \cite{linuxnamespaces} to create isolated environments on a host system \cite{merkel2014docker}. These environments function as lightweight virtual machines, excluding emulation software but including essential applications and runtimes as files. Docker \cite{docker} tandardizes container runtime and image specifications, defining how containers execute and how image manifests store metadata for application launches. Docker enhances container functionality by enabling portable deployment, versioning, and component reuse through binary objects called \emph{blobs}. These blobs encapsulate applications, allowing seamless transfer and rapid installation on Docker-enabled systems.

\noindent \textbf{Virtualization services.}
Figure \ref{fig:bck_docker} illustrates the Docker stack used in our system. Users interact with the Docker daemon (\texttt{dockerd}) via the command-line interface (\texttt{docker-cli}). \texttt{dockerd} manages features such as container volume handling and Ethernet communication through an HTTP REST API, while establishing a virtual network for communication between users and containers. Below this layer, the container engine (\texttt{containerd}) manages images and containers, and the low-level runtime (\texttt{runc}) configures namespaces, cgroups, and initiates containers.

Figure \ref{fig:bck_image} shows the container creation process from an image. Docker retrieves a blob from a user-defined location (\mbox{\ding{182}}), which \texttt{containerd} unpacks based on image specifications (\mbox{\ding{183}}), generating a configuration file and image layers. The layers are merged into a read-only \emph{lower directory} using a Docker storage driver (e.g., overlay). This \texttt{runc} then creates a writable \emph{upper directory} and merges it with the lower directory to form a root filesystem (\texttt{rootfs}) (\ding{184}). \texttt{rootfs} with runtime configurations enables independent application execution across varied environments (\ding{185}).

Note that cgroups and namespaces are OS kernel features, not intrinsic to containers. While containers utilize these features for multi-tenancy-throttling requests with cgroups and isolating resources with namespaces, their execution environment is determined by runtime and image specifications. Docker's stack is crucial for mounting and executing containers, as each container operates as a standard process on a Docker-enabled system.
\vspace{-8pt}

\begin{figure}
  \centering
  \includegraphics[width=\linewidth]{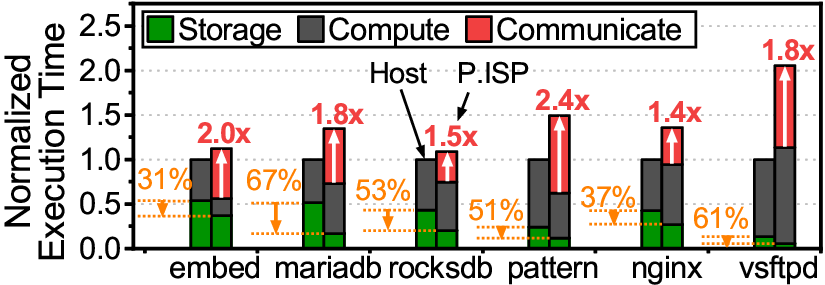}

  \vspace{10pt}
  \caption{Performance impact analysis.} \label{fig:motiv}
\end{figure}
%-------------------------------------------------------------------------------

%-------------------------------------------------------------------------------
\section{CHALLENGES IN ISP}
\label{sec:challenge}
Various ISP models have been proposed to offload operations to storage and thereby reduce data-movement overhead. For example, Summarizer \cite{koo2017summarizer} and KAML \cite{jin2017kaml}, representative examples in this field, respectively handle data filtering and key-value database operations by offloading specific kernels to the storage side. Willow \cite{seshadri2014willow} and Biscuit \cite{gu2016biscuit} address broader user requirements, allowing users to compile and offload kernels dynamically via firmware APIs.

Despite these and other efforts, such ISP models have not seen wide adoption. One major reason is that existing models often require static kernels or vendor-specific APIs for offloading, significantly reducing user convenience. Specifically, five key challenges must be overcome to improve these limitations: manual ISP implementation, disregard for file layout, kernel context switching, device reliance, and data vulnerability. Addressing these challenges demands an ISP model that supports execution and environment independence, and flexible virtualization for diverse applications.

\subsection{Performance Impact Assessment}

We evaluate a revised iteration of the Programmable-ISP model (\texttt{P.ISP}) \cite{seshadri2014willow, gu2016biscuit}. Detailed workload characteristics and evaluation environments are discussed in the \quotes{EVALUATION} section. Figure \ref{fig:motiv} illustrates a breakdown of ISP execution times into three components: ISP computation latency (\texttt{Compute}), storage backend delay (\texttt{Storage}), and host-to-ISP communication/synchronization overhead (\texttt{Communicate}). For comparison, we also evaluate a host-only system (\texttt{Host}). Our findings reveal that \texttt{Storage} accounts for 38\% of the total execution time, highlighting the potential for reducing application latency. By processing data in storage, \texttt{P.ISP} reduces \texttt{Storage} latency by 50\% compared to \texttt{Host}, effectively mitigating data movement overhead. However, this improvement is offset by a 1.4$\times$ increase in overall end-to-end latency, primarily due to \texttt{Communicate}, which constitutes 43\% of \texttt{P.ISP} latency.

Addressing these challenges requires making ISP models host-independent and enabling autonomous execution. We believe that, by eliminating communication overhead, ISP can achieve better performance. We also believe that virtualizing the ISP model can enable it to function as a secure, portable sandbox, paving the way for broader ISP adoption.
%-------------------------------------------------------------------------------

%-------------------------------------------------------------------------------
\section{ISP CONTAINERIZATION}
\label{sec:overview}
\begin{figure}
  \includegraphics[width=\linewidth, bb=0 15 256 70]{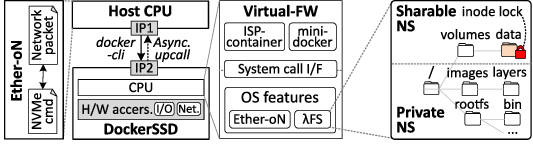}
  \vspace{0pt}
  \begin{subfigure}{0.69\linewidth}
    \centering
    % \vspace{-12pt}
    \caption{ISP containerization.} \label{fig:isp}
  \end{subfigure}
  \begin{subfigure}{0.30\linewidth}
    \centering
    % \vspace{-12pt}
    \caption{Media mgmt.} \label{fig:backmedia2}
  \end{subfigure}

  \vspace{18pt}
  \caption{High-level view of DockerSSD.}
  \vspace{3pt}
  \label{fig:high-level-view}
\end{figure}

Unlike conventional ISPs that rely on static kernels or tasks, DockerSSD establishes virtual ISP environment, enabling the host to process data near flash using containerized applications, called \emph{ISP-containers}. This section explores how DockerSSD achieves ISP containerization through firmware, addressing challenges like manual ISP implementation, context switches, and data vulnerability. By managing backend media to reduce simultaneous access and leveraging firmware-level virtualization, DockerSSD eliminates device reliance and file layout limitations, ensuring containers operate independently across diverse systems.

\vspace{-8pt}
\subsection{High-level Overview of DockerSSD}
\label{subsec:overview}

To implement containerization in storage, DockerSSD addresses three technical challenges. First, it requires a new communication interface compatible with both existing storage and network stacks for effective ISP-container management. Second, the ISP model must ensure consistency across a computing-enabled storage pool, enabling independent operation of all components. Third, the underlying firmware must recognize container structures and execute them with minimal overhead.

Figure \ref{fig:isp} illustrates the architecture of DockerSSD. To overcome these challenges, it integrates two key components: i) a system driver introducing a novel ISP management interface, \emph{Ether-oN}, and ii) a firmware environment equipped with Docker stack, \emph{Virtual-FW}. \emph{Ether-oN} facilitates socket-based Ethernet communication over PCIe by virtualizing NVMe. It achieves this through a driver supporting network-to-storage packet translation and asynchronous upcalls. 
By assigning individual IP addresses to each endpoint, Ether-oN enables users to issue ISP-related requests to Virtual-FW via \texttt{docker-cli}. 
Details about Ether-oN are provided in the "ETHERNET OVER NVME" section.
\emph{Virtual-FW} incorporates minimal OS features into SSD, emulating system call interfaces on bare-metal hardware. It includes \emph{mini-docker}, which implements core functions of Docker stack, enabling firmware-level ISP containerization.
Further details about Virtual-FW are provided in the "DOCKER-ENABLED FIRMWARE" section.

In this study, we demonstrate how Virtual-FW and Ether-oN allow users to run full-scale applications near flash using familiar Docker commands, without relying on vendor-specific interfaces. 
Furthermore, Ether-oN supports all essential Ethernet functionalities, allowing storage components to operate as independent nodes within a computing-enabled storage pool, communicating with the host seamlessly and autonomously.

\begin{figure}
  \centering
  % \vspace{-2pt}
  \includegraphics[width=1\linewidth]{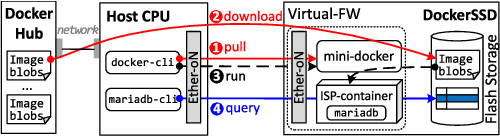}
  \vspace{5pt}
  \caption{Application execution example.} \label{fig:app-example}
  \vspace{0pt}
\end{figure}

Figure \ref{fig:app-example} illustrates this process. Users initiate a blob download from Virtual-FW using \texttt{docker pull} (\red{\circleNum{1}}, \red{\circleNum{2}}) and launch the corresponding ISP-container via \texttt{docker run}, all without requiring application-level modifications (\circleNum{3}). During ISP-container execution, users can submit non-I/O, ISP-specific requests through Ether-oN (\blue{\circleNum{4}}). For example, a query submitted via \texttt{mariadb-cli} is processed over Ethernet. 
While Virtual-FW is optimized to minimize the execution cost of ISP-containers through efficient system call emulation and firmware-level functionality, ISP-containers themselves can generate I/O and network requests for data processing and communication as needed.

\vspace{5pt}

\vspace{-5pt}
\subsection{Backend Media Management}
\label{subsec:lfs}
DockerSSD introduces the \emph{Lambda filesystem} ($\lambda$FS) for efficient backend flash management, as shown in Figure \ref{fig:backmedia2}. 
To enabled seamless sharing of files and directories between the host and storage, $\lambda$FS integrates with the host's file structures, based on EXT4, while ensuring protection against unauthorized access by ISP-containers.
To handle file concurrency and address potential vulnerabilities, $\lambda$FS partitions the media into two NVMe namespaces, supported by the NVMe subsystem: 
i) the private namespace (\emph{private-NS}) and ii) the sharable namespace (\emph{sharable-NS}). 
The private-NS is isolated from the host, while the sharable-NS is accessible to both the host and ISP-containers. 

In this configuration, the NVMe subsystem, managed by HIL, exposes two PCIe functions, resembling storage-side ports.
One function is associated with Virtual-FW, encompassing both private- and sharable-NS, while the other is linked to the host and includes only the sharable-NS.
This design ensures secure and efficient file sharing and isolation across storage and ISP-containers.
$\lambda$FS allocates private-NS for container and OS-level virtualization runtimes. Thus, ISP-related contents needing protection, such as image layers (\texttt{/images/}) and container data (\texttt{/rootfs/}), are kept invisible to the users. Conversely, the sharable-NS is used for data that the host needs to place/retrieve and ISP-container processes in storage.
This allows ISP-containers and users to share necessary in-out data, but it requires a lock for the data across all I/Os.

% To tackle this, $\lambda$FS employs a new in-memory data structure, the \texttt{inode} lock. This lock synchronizes with the host's \texttt{inode} cache via Ether-oN when an ISP-container binds a host FS directory or file to $\lambda$FS. The file is accessible only if the \texttt{inode} reference count is zero, resolving concurrent access issues. When the ISP-container gets file access permission, VFS invalidates its \texttt{inode} cache, referring to the storage's latest information. Note that when the host and an ISP-container access the same partition concurrently, the host may retain a stale \texttt{inode} in the VFS. This can lead to filesystem corruption or render files/directories updated by the ISP-container inaccessible. To address this, DockerSSD synchronizes updates between the host and ISP-container, even without host involvement.

To tackle this, $\lambda$FS employs a new in-memory data structure, the \texttt{inode} lock. 
This lock synchronizes with the host's \texttt{inode} cache via Ether-oN when an ISP-container binds a host FS directory or file to $\lambda$FS for data processing.
For synchronization, $\lambda$FS adds a reference counter to the inode of the \emph{virtual file system} (VFS) on the host, protecting it with the existing inode's semaphore. This counter updates when the target file (or its directory file) is opened or closed. VFS and $\lambda$FS then send a special packet
via Ether-oN to update it. The file is accessible only if the \texttt{inode} reference counter is zero, resolving concurrent
access issues. When the ISP-container gets file access permission, VFS invalidates its \texttt{inode} cache, referring to the storage's latest information. 
Note that when both the host and an ISP-container access the same partition concurrently, the host may retain a stale \texttt{inode} in the VFS. This can lead to filesystem corruption or render files and directories updated by the ISP-container inaccessible. 
To address this, DockerSSD must synchronize updates between the host and ISP-container, even without direct host involvement. 
The locking mechanism implemented in DockerSSD is designed exclusively for synchronization and does not require persistence. 
In the event of a power failure, the lock is not retained, as the host can restore the filesystem and restart the ISP-container from its initial state.

\begin{figure}
  \centering
  \includegraphics[width=1\linewidth]{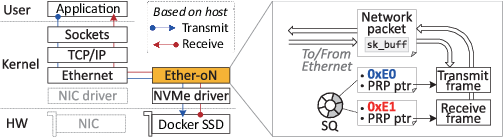}
  \begin{subfigure}{\linewidth}
    \centering
    \begin{tabularx}{\textwidth}{
      p{\dimexpr.45\linewidth-2\tabcolsep-1.3333\arrayrulewidth}% column 1
      p{\dimexpr.55\linewidth-2\tabcolsep-1.3333\arrayrulewidth}% column 2
      }
      \vspace{-3pt} \caption{Software stack.} \label{fig:ether_on1} &
      \vspace{-3pt} \caption{Transmit/receive frames.} \label{fig:ether_on2}
    \end{tabularx}
  \end{subfigure}
  \vspace{2pt}
  \caption{Ether-oN Interface.} \label{fig:ether_on}
\end{figure}

\vspace{-5pt}
\section{ETHERNET OVER NVME}
\label{subsec:ether-on}
The Docker stack and container interface rely on Ethernet, making it a natural choice for ISP-related service communication. 
To enable this, we overlay standard socket-based networking onto NVMe protocol (Ether-oN). 
Figure \ref{fig:ether_on1} illustrates Ether-oN's architecture. Host applications communicate with DockerSSD using a conventional network stack, interacting with Virtual-FW via Ethernet protocols such as TCP/IP. 

Ether-oN establishes an intranet between the host and DockerSSD, eliminating the need for source-level modifications in host applications. 
Its kernel driver creates a virtual network adapter on the host, directly linking to DockerSSD. 
This adapter integrates with the host's network stack to translate Ethernet packets into NVMe commands,
which are sent to DockerSSD for processing. This approach
introduces two technical challenges. First, unlike traditional
network environments, NVMe lacks a mechanism to handle
in-bound requests, as PCIe devices typically cannot issue
NVMe commands to the host. ISP-containers, however,
require such an upcall mechanism for user communication.
Second, host users need seamless access to DockerSSD
through both the network and storage stacks for ISP and
block I/O services. To address these issues, we introduce
two NVMe vendor-specific commands, explained shortly.

\noindent \textbf{Network support using NVMe.}
The two new NVMe commands to support Ethernet over NVMe are i) \emph{transmit} and ii) \emph{receive} frames. 
They share the same structure as standard NVMe commands but utilize vendor-specific operation codes (0xE0 - 0xE1) reserved for custom implementations (cf. Figure \ref{fig:ether_on2}). 
When the Ether-oN driver receives an Ethernet request, it extracts the network packet buffer (\texttt{struct sk\_buff}) from the frame and allocates a 4KB-aligned kernel page. 
The \texttt{sk\_buff}, including headers, payloads, and checksums, is copied to the page.
The Ether-oN driver then creates an NVMe command, populates its PRP field with the kernel page's address, sets the operation code to transmit, and submits the command to the driver.

\noindent \textbf{Enabling inbound network services.}
To allow DockerSSD to send service requests to the host (e.g., network responses
or remote access requests), we implement an asynchronous
upcall mechanism using pre-allocated NVMe commands.
During kernel initialization, Ether-oN pre-submits a set of
NVMe commands, each configured as a receive frame, to
the NVMe SQ. For every pre-allocated command, Ether-oN
assigns a kernel page and inserts a reception code. 
DockerSSD holds these commands until an ISP-container sends an Ethernet frame to the host. At that point, DockerSSD copies the \texttt{sk\_buff} to the corresponding kernel page and completes the outstanding command. Upon completion,
Ether-oN translates the NVMe command into an Ethernet frame and delivers it to the network stack. To maintain
communication, Ether-oN immediately submits a new receive
frame to DockerSSD. The NVMe interface supports
as many upcall requests as the SQ allows. Based on our
preliminary studies, we use four pre-allocated commands
per SQ to balance efficiency and resource utilization.

% Ether-oN establishes an intranet between the host and DockerSSD, translating Ethernet packets into NVMe commands sent to DockerSSD. This introduces two challenges. 
% First, NVMe lacks a in-bound request handling mechanism. Second, host users need seamless access to DockerSSD via network and storage stacks for ISP and block I/O services. To address these issues, we introduce two NVMe vendor-specific commands, transmit and receive frames. 
% When Ether-oN driver receives an Ethernet request, it extracts the packet buffer, allocates a kernel page, copies the buffer to the kernel page. The Ether-oN driver then creates an NVMe command with the kernel page, sets the operation code to transmit, and submits the command to the driver.

% To allow DockerSSD to request host services, we implement an asynchronous upcall mechanism using pre-allocated NVMe commands. During kernel initialization, Ether-oN pre-submits a set of NVMe commands, each configured as a receive frame, to the NVMe SQ. For every pre-allocated command, Ether-oN assigns a kernel page and inserts a reception code. DockerSSD holds these commands until an ISP-container sends an Ethernet frame to the host. At that point, DockerSSD copies the frame's buffer to the corresponding kernel page and completes the outstanding command. Upon completion, Ether-oN translates the NVMe command into an Ethernet frame and delivers it to the network stack. To maintain communication, Ether-oN immediately submits a new receive frame to DockerSSD.

\vspace{-6pt}
\section{DOCKER-ENABLED FIRMWARE}
\label{subsec:virtualfw}

\begin{figure}
  \includegraphics[width=\linewidth, bb=0 10 242 66]{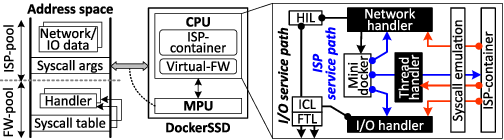}
  \begin{subfigure}{0.49\linewidth}
    \centering
    \vspace{-10pt}
    \caption{Memory mgmt.} \label{fig:memory-mgmt}
  \end{subfigure}
  \begin{subfigure}{0.49\linewidth}
    \centering
    \vspace{-10pt}
    \caption{Service paths.} \label{fig:service-path}
  \end{subfigure}
  \vspace{10pt}
  \caption{Internal organization of our Virtual-FW.}
  \label{fig:internal-org}
\end{figure}

While Ether-oN enables network-based communication on
block devices, existing firmware lacks the flexibility required
for Ethernet-based ISP computing and resource management.
To address this, Virtual-FW integrates essential
OS features and a container environment into the current
I/O service path (HIL$\Rightarrow$ICL$\Rightarrow$FTL).

\noindent \textbf{Integration of OS Features.}
Figure \ref{fig:internal-org} illustrates how the
ISP service path integrates into the existing firmware-level
I/O service path. DockerSSD implements three handlers
for OS features: \emph{thread}, \emph{I/O}, and \emph{network} management,
positioned between HIL and ICL. The \emph{thread handler} manages its bare-metal DRAM in page-granular partitions:
the \emph{FW-pool} and \emph{ISP-pool}. 
The FW-pool contains tables for all three handlers using system call emulation, while
the ISP-pool handles call arguments and other data types.
Execution is distinguished by DockerSSD's CPU modes,
with privileged mode required for FW-pool access, enforced
by the memory protection unit. This safeguards Virtual-FW
while eliminating the need for data copying between pools,
as privileged mode allows Virtual-FW to access the ISP pool
directly, avoiding mode-switching overhead.

The \emph{I/O handler} processes only I/Os generated by ISP containers
for data stored in DockerSSD, without requiring heavy block management layers like the block layer or
NVMe software stack. It implements $\lambda$FS (see "Backend Media Management") and exposes it through Virtual-FW's emulation interfaces. 
Key functionalities include \emph{path walking}, which maps logical block addresses (LBAs) to filenames, and \emph{I/O node caching}, which caches these mappings
for faster access. 
The \emph{network handler} manages network communication through TCP/IP features, including establishing communication channels and handling network packets. It employs a TCP finite state machine to track
socket communication states and performs packet encapsulation
and parsing for the channel management.

\begin{figure}
  % \vspace{-5pt}
  \begin{minipage}[t]{.59\linewidth}
    \resizebox{\linewidth}{!}{%
    \setlength\tabcolsep{3pt}
    \begin{tabular}{|c|l|l|}
    \hline
    \begin{tabular}[c]{@{}c@{}}\textbf{Handler}\\\textbf{(\# syscall)}\end{tabular}
                          & \multicolumn{1}{c|}{\textbf{Category}}
                                                      & \multicolumn{1}{c|}{\textbf{Example}}
                                                                                \\ \hline
    \multirow{4}{*}{\begin{tabular}[c]{@{}c@{}}Thread\\ Handler\\ (65)\end{tabular}}
                          & Process management   & \texttt{fork exit}      \\ \cline{2-3}
                          & Memory management    & \texttt{brk mmap}       \\ \cline{2-3}
                          & Inter-Process comm.  & \texttt{pipe mq\_open}  \\ \cline{2-3}
                          & Lock \& signal mgmt. & \texttt{futex}          \\ \hline
    \multirow{3}{*}{\begin{tabular}[c]{@{}c@{}}I/O\\ Handler\\ (43)\end{tabular}}
                          & File/dir mgmt.       & \texttt{openat mkdir}   \\ \cline{2-3}
                          & File I/O \& link     & \texttt{read symlink}   \\ \cline{2-3}
                          & Permission           & \texttt{chmod chown}    \\ \hline
    \multirow{3}{*}{\begin{tabular}[c]{@{}c@{}}Network\\ Handler\\ (25)\end{tabular}}
                          & Polling APIs         & \texttt{epoll\_create}  \\ \cline{2-3}
                          & Socket APIs          & \texttt{socket bind}    \\ \cline{2-3}
                          & Network comm.        & \texttt{sendto}         \\ \hline
    \end{tabular}
    }
  \end{minipage}
  \begin{minipage}[t]{.397\linewidth}
    \resizebox{\linewidth}{!}{%
    \setlength\tabcolsep{3pt}
    \begin{tabular}{|c|l|}
    \hline
    \textbf{Category} & \multicolumn{1}{c|}{\textbf{CLI command}}  \\ \hline
    \multirow{2}{*}{\begin{tabular}[c]{@{}c@{}}Image\\ management\end{tabular}}
      & \texttt{docker rmi}      \\
      & \texttt{docker pull}     \\ \hline
    \multirow{6}{*}[-6pt]{\begin{tabular}[c]{@{}c@{}}Container\\ life cycle\\ management\end{tabular}}
      & \texttt{docker create}   \\
      & \texttt{docker run}      \\
      & \texttt{docker start}    \\
      & \texttt{docker stop}     \\
      & \texttt{docker restart}  \\
      & \texttt{docker kill}     \\
      & \texttt{docker rm}       \\ \hline
    \multirow{2}{*}{\begin{tabular}[c]{@{}c@{}}Container\\ monitoring\end{tabular}}
      & \texttt{docker logs}     \\
      & \texttt{docker ps}       \\ \hline
    \end{tabular}
  }
  \end{minipage}
  \captionsetup{type=table}
  \begin{subfigure}{\linewidth}
    \centering
    \begin{tabularx}{\textwidth}{
      p{\dimexpr.58\linewidth-2\tabcolsep-1.3333\arrayrulewidth}% column 2
      p{\dimexpr.45\linewidth-2\tabcolsep-1.3333\arrayrulewidth}% column 1
      }
      \vspace{-7pt} \caption{System calls.} \label{tab:syscall} &
      \vspace{-7pt} \caption{Docker commands.} \label{tab:docker}
    \end{tabularx}
  \end{subfigure}
  \vspace{3pt}
  \caption{Virtual-FW's OS features and commands.} \label{tab:lambdastore}
  \vspace{0pt}
\end{figure}

\noindent \textbf{System call emulation.}
Virtual-FW emulates system calls to enable seamless container execution within storage, eliminating
the need for application-level changes or recompilation.
To accommodate the SSD's limited computing resources, the emulation is implemented using lightweight
function wrappers. Table \ref{tab:syscall} summarizes the types and quantities of system calls emulated, along with the corresponding handler responsible for each. 
This approach minimizes memory overhead by removing unnecessary system
function overrides from the call path. For example, \texttt{glibc}'s \texttt{open} internally invokes \texttt{openat}, which is designed for
compatibility across CPU architectures but is redundant for Virtual-FW. 
In terms of latency, it significantly reduces
context switch overhead, lowering the impact of running
containers on SSDs. Unlike a fully-fledged OS, which
perform context switches when system calls return to userland,
Virtual-FW's function-level implementation avoids such boundaries, further optimizing call performance.

\begin{figure*}[t]
  \centering
  \includegraphics[width=1\linewidth]{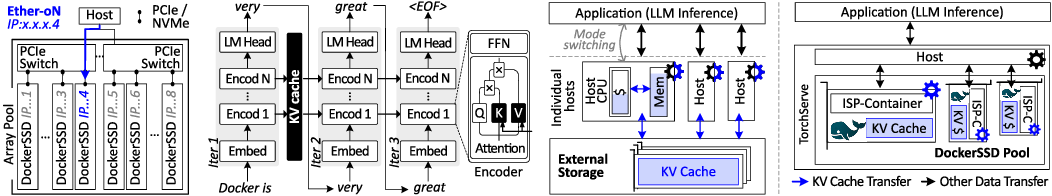}

  \vspace{3pt}
  \begin{subfigure}{1\linewidth}
      \centering
      \renewcommand*{\arraystretch}{0.3}
      \begin{tabularx}{\textwidth}{
          p{\dimexpr.19\linewidth-2\tabcolsep-1.3333\arrayrulewidth}
          p{\dimexpr.31\linewidth-2\tabcolsep-1.3333\arrayrulewidth}
          p{\dimexpr.24\linewidth-2\tabcolsep-1.3333\arrayrulewidth}
          p{\dimexpr.24\linewidth-2\tabcolsep-1.3333\arrayrulewidth}
          }
            \vspace{3pt}\caption{Disaggregation.} \label{fig:docker_pool}
          & \vspace{3pt}\caption{LLM architecture.} \label{fig:llm_arch}
          & \vspace{3pt}\caption{Individual hosts.} \label{fig:inference_host}
          & \vspace{3pt}\caption{DockerSSD pool.} \label{fig:inference_dockerssd}
      \end{tabularx}
      \vspace{1pt}
  \end{subfigure}
  \vspace{-9pt}
	\caption{DockerSSD disaggregation and LLM distributed inference.} \label{fig:dockerssd_llm}
  \vspace{5pt}
\end{figure*}

\noindent \textbf{Firmware-level container environment.}
The container environment within DockerSSD focuses on storing, running, and managing containers near flash memory, rather than supporting the full range of Docker functionalities. To achieve this, Virtual-FW introduces \emph{mini-docker}, a streamlined implementation that supports 11 essential Docker commands (out of 106), as summarized in Table \ref{tab:docker}. It establishes a container environment linked to \texttt{dockerd}, \texttt{containerd}, and \texttt{runc}.
Similar to \texttt{dockerd}, mini-docker communicates with the host's \texttt{docker-cli} using HTTP. To manage ISP-containers, a user issues an HTTP command specifying DockerSSD's IP address and the image name to execute. Mini-docker parses the command and performs the corresponding action. For example, when an image management command (\texttt{docker pull}/\texttt{rmi}) is issued, it stores the image blob and its manifest in $\lambda$FS (under \texttt{/images/blobs} and \texttt{/images/manifest}).

The manifest includes details about the target application, such as its entry script and required image layers for \texttt{rootfs}. When the host issues commands to create or start ISP (\texttt{docker create/start}), mini-docker invokes the thread handler to generate an ISP-container, using the entry script (similar to \texttt{containerd}). 
It then mounts the \texttt{rootfs} to the ISP-container. To monitor running ISP-containers, mini-docker logs information (e.g., \texttt{stdout} and \texttt{stderr}) to $\lambda$FS under \texttt{/containers/<id>/rootfs/log}. The logs can be transferred to the host via Virtual-FW, enabling real-time analysis.

%-------------------------------------------------------------------------------

%-------------------------------------------------------------------------------
\section{RESOURCE DISAGGREGATION}
\label{sec:swfw}

With Ether-oN and Virtual-FW, DockerSSD gains its IP address and can run containerized applications as an independent node. This enables the creation of a computing-enabled storage pool by disaggregating DockerSSDs from their dedicated hosts. As illustrated in Figure \ref{fig:docker_pool}, DockerSSDs can form an array pool connected via one or more PCIe switches. Multiple arrays can be integrated into a cluster using a switch tray or chassis unit for larger-scale deployments. Despite the PCIe connectivity, Ether-oN assigns each DockerSSD a unique IP address by converting Ethernet protocols to NVMe (and vice versa). This allows the host to assign specific applications to any node in the array or cluster and monitor them through mini-docker logs.

\noindent \textbf{Computing-enabled storage pool.}
There are two primary methods for offloading computing to the computing-enabled storage pool. The first involves offloading independent applications across DockerSSDs in the pool, allowing each node to operate autonomously. The second, preferred in this work, groups multiple DockerSSDs into a distributed computing system. For distributed computing, DockerSSDs leverage frameworks such as \texttt{docker-compose} \cite{dockercompose} or Kubernetes \cite{kubernetes} to orchestrate containers across nodes efficiently.

As an use-case for our computing-enabled storage pool, we set up a distributed inference system for a large-scale pre-trained model. Specifically, we implement a LLM, well-suited to this scenario. Due to memory and computational limitations, single nodes or devices cannot manage inference tasks for trillion-scale model parameters. As model size increases, LLMs employ distributed inference techniques, leveraging methods such as pipeline parallelism, data parallelism, and tensor parallelism. In pipeline parallelism, DockerSSDs allocate different layers of the model across multiple ISP devices. For data parallelism and tensor parallelism, the model divides data either based on device awareness within the storage pool or along dimensionality, ensuring efficient utilization of the distributed infrastructure. TorchServe model serving platform can help our distributed inference test, which utilizes Kubernetes for container execution.

\noindent \textbf{Distributed inference with DockerSSDs.}
As shown in Figure \ref{fig:llm_arch}, LLMs generate the next token based on a given set of tokens. Each iteration reuses the tokens calculated by far as input for the next step, progressively expanding the sequence as the inference continues. LLMs consist of multiple stacked encoders, each composed of an attention layer and a feed-forward network (FFN) layer. The attention layer calculates the importance of each token relative to others using the key (K) vector, which represents the significance of the token, and the value (V) vector, which encodes the actual semantic information of the token. Because LLMs reuse tokens during each inference iteration, recalculating the K and V vectors for the same tokens in every iteration introduces unnecessary computational overhead. To address this, modern LLMs implement a \emph{key-value} (KV) caching, which stores the computed K and V vectors in memory for reuse in subsequent iterations, significantly reducing computational burden.

\begin{figure}
  \vspace{-4pt}
  \centering
  \begin{minipage}[t]{.73\linewidth}
    \centering
    \includegraphics[width=\linewidth]{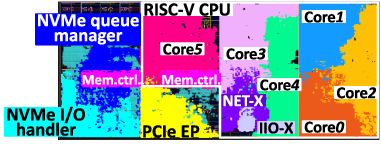}
\end{minipage}
\begin{minipage}[t]{.25\linewidth}
    \includegraphics[width=\linewidth]{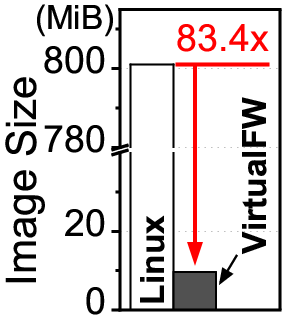}
\end{minipage}
\vspace{4pt}

  \begin{minipage}[b]{.73\linewidth}
      \vspace{-8pt}
      \captionsetup{labelsep=newline, justification=raggedright, singlelinecheck=false}
      \caption{DockerSSD implementation.}
      \label{fig:eval_floorplan}
  \end{minipage}
  \begin{minipage}[b]{.25\linewidth}
      \vspace{-8pt}
      \caption{\\ Image size.}
      \label{fig:eval_virtualfw}
  \end{minipage}
  \vspace{-12pt}
\end{figure}

Figures \ref{fig:inference_host} and \ref{fig:inference_dockerssd} compare LLM inference performance when served from individual hosts versus our computing-enabled storage pool in a distributed manner, respectively. KV caching makes LLM inference a memory-intensive application, often requiring additional memory capacity through external storage. In distributed environments, this can lead to cache pollution, frequent mode switches, and unnecessary data copies on the hosts. In contrast, each ISP device in our storage pool manages KV caching directly on its flash storage, avoiding these inefficiencies. In addition, distributed inferences handled by TorchServe utilize the storage pool effectively, enabling the processing of much longer token sequences and providing greater context for each service. This approach enhances the overall capacity and scalability of the LLM system.
%-------------------------------------------------------------------------------

%-------------------------------------------------------------------------------
\section{EVALUATION}
\label{sec:evaluation}
\noindent \textbf{Prototype and methodology.}
We built a DockerSSD prototype using a 16nm FinFET FPGA \cite{virtexup}. The NVMe hardware IPs \cite{jung2020openexpress} were adapted to include a multi-core processor, integrating RTL into DockerSSD's frontend. 
As shown in Figure \ref{fig:eval_floorplan}, the prototype includes a queue manager, I/O engine, and PCIe controller for NVMe tasks. 
Six RISC-V in-order cores run Virtual-FW, connected via AXI/TileLink buses. 
For the backend, two DDR4 controllers emulate flash using a modified multi-channel timing model \cite{jung2017simplessd}.
We modified 3.4K LOC for Ether-oN/$\lambda$FS and 7.5K LOC for firmware and storage modules (EXT4-based). 
Figure \ref{fig:eval_virtualfw} shows Virtual-FW reduced the Linux binary size by 83.4$\times$, making it suitable for embedded processors.
Since no single device can accommodate all hardware/software configurations or ISP communication methods, we evaluated various ISP models using the gem5 full-system simulator \cite{binkert2011gem5}. 
The simulation framework has been cross-validated with our hardware RTL and synthesized prototype backend, and integrated into a cycle-accurate SSD simulator \cite{jung2017simplessd}.
The simulator, validated against the prototype's backend, integrates a cycle-accurate SSD simulator and power model. 
The host has a 3.8GHz CPU and 64GB DDR4, while storage includes an NVMe SSD with 48 MLC flashes across 12 channels. The frontend uses a 2.2GHz processor and 2GB DRAM, comparable to prior studies.

\noindent \textbf{Data processing models.}
We designed six data processing models for evaluation. The \texttt{Host} model, representing a baseline non-ISP setup, serves as the reference.
Two programmable ISP models, \texttt{P.ISP-R} and \texttt{P.ISP-V}, were implemented using offloading methods from prior studies \cite{seshadri2014willow, gu2016biscuit}. \texttt{P.ISP-R} uses as its interface \cite{seshadri2014willow}, while \texttt{P.ISP-V} employs NVMe vendor-specific commands to minimize interface overhead \cite{gu2016biscuit}. 
Both models assume that most ISP functions are offloaded, except for system-specific modules like file I/O management, which rely on the host's runtime and file systems. In our setup, \texttt{P.ISP-R/V} only accesses LBAs when ISP kernels require a new file, reducing overhead.
We also evaluated three DockerSSD variations. \texttt{D-Naive} uses separate processor and controller complexes, with one running an ISP-container and the other running firmware with a full Linux \cite{shadley2018deployment}. 
\texttt{D-FullOS} modifies this by integrating DockerSSD's hardware, enabling the ISP-container and firmware to run on the same complex. \texttt{D-VirtFW} replaces Linux in \texttt{D-FullOS} with Virtual-FW for lightweight operation.

\begin{table}[t]
  \vspace{-2pt}
  \centering
  \setlength\tabcolsep{3pt}
  \setlength\extrarowheight{1pt}
  \resizebox{1.01\linewidth}{!}{%
    \begin{tabular}{|cl|rrrrrrr|}
    \hline
    %\noalign{\vskip 0pt}
    \setlength{\baselineskip}{4pt}
    \textbf{Program}      & \multicolumn{1}{c|}{\textbf{Workload}}
                          & \multicolumn{1}{c}{\textbf{\begin{tabular}[c]{@{}c@{}}I/O\\ size\end{tabular}}}
                          & \multicolumn{1}{c}{\textbf{\begin{tabular}[c]{@{}c@{}}I/O\\ count\end{tabular}}}
                          & \multicolumn{1}{c}{\textbf{\begin{tabular}[c]{@{}c@{}}\# sys-\\ calls\end{tabular}}}
                          & \multicolumn{1}{c}{\textbf{\begin{tabular}[c]{@{}c@{}}\# path\\ walk\end{tabular}}}
                          & \multicolumn{1}{c}{\textbf{\begin{tabular}[c]{@{}c@{}}\# files\\ opened\end{tabular}}}
                          & \multicolumn{1}{c}{\textbf{\begin{tabular}[c]{@{}c@{}}\# TCP\\ packet\end{tabular}}}
                          & \multicolumn{1}{c|}{\textbf{\begin{tabular}[c]{@{}c@{}}Exec.\\ Time\end{tabular}}} \\ \hline
                          %\setlength{\baselineskip}{8pt}
                          %\vspace{-0.5em}
                          \multirow{2}{*}{\begin{tabular}[c]{@{}c@{}}embed- \\ \end{tabular}}
                          &      rm1               &  1.3GB &  317K &  1.3M &   9K &    260 &    0 &   8s \\ % \cline{2-7}
                          &      rm2  &  5.8GB &  1.4M &  1.7M &   9K &    320 &    0 &  24s \\ \hline
    \multirow{2}{*}{\begin{tabular}[c]{@{}c@{}}mariadb- \\ \end{tabular}}
                          &    tpch4                 & 17.1GB &  1.1M &  1.1M &  37K &    250 &  160 &  25s \\ % \cline{2-7}
                          &   tpch11                 &  6.2GB &  400K &  361K &  38K &    260 &  190 &   8s \\ \hline
    \multirow{2}{*}{\begin{tabular}[c]{@{}c@{}}rocksdb- \\ \end{tabular}}
                          &     read              &  4.1GB &  431K &  1.1M &   9K &   1.2K &    0 &  14s \\ % \cline{2-7}
                          &    write              & 18.5GB &   24K &  285K &   9K &   3.6K &    0 &  24s \\ \hline
    \multirow{3}{*}{\begin{tabular}[c]{@{}c@{}}pattern- \\ \end{tabular}}
                          &     find             &  2.4GB &  381K &  1.8M & 359K &   352K &    0 &  11s \\ % \cline{2-7}
                          &     line             &  1.7GB &  262K &  1.7M & 476K &   235K &    0 &  11s \\ % \cline{2-7}
                          &     word             &  2.1GB &  340K &  2.2M & 618K &   307K &    0 &  10s \\ \hline
    \multirow{3}{*}{\begin{tabular}[c]{@{}c@{}}nginx-\end{tabular}}
                          &     web0              &  7.5GB &  126K &  665K & 126K &   4.4K & 543M &   9s \\ % \cline{2-7}
                          &     web1              &  0.9GB &   50K &  344K & 109K &     2K & 154K &   3s \\ % \cline{2-7}
                          & filedown \            & 13.5GB &  109K &   30K &   1K &     40 & 155K &   6s \\ \hline
    vsftpd-
                          &   fileup           & 12.1GB &   93K &  5.4M & 127K &   115K & 1.2M &   2s \\ \hline
    \end{tabular}
  }
  \vspace{3pt}
  \caption{Workload characteristics.} \label{tbl:eval_workload}
  \vspace{0pt}
\end{table}

\begin{figure*}[]
  \centering
  \includegraphics[width=1\linewidth]{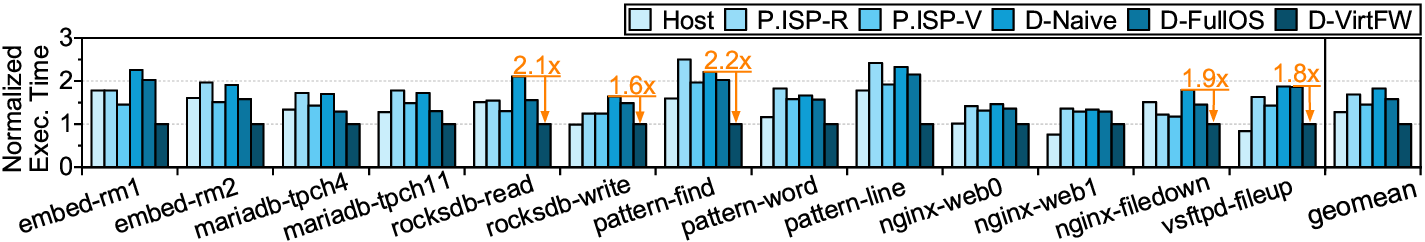}
  \vspace{-3pt}
  \caption{Overall performance analysis.} \label{fig:eval_latency}
\end{figure*}

\begin{figure*}
  \centering
  \vspace{20pt}
  \includegraphics[width=1\linewidth]{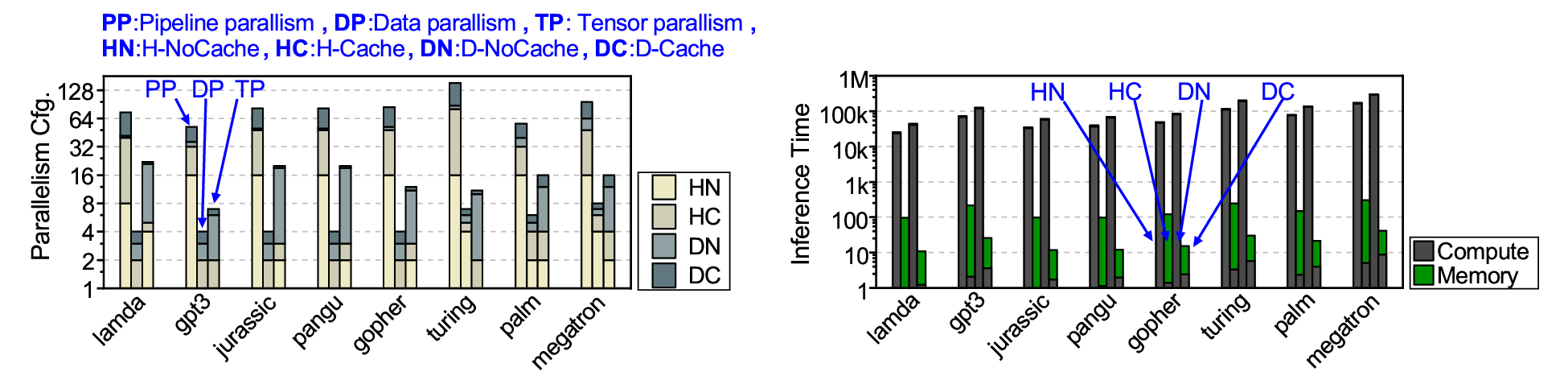}
  \begin{subfigure}{1\linewidth}
      \centering
      \renewcommand*{\arraystretch}{0.3}
      \begin{tabularx}{\textwidth}{
          p{\dimexpr.5\linewidth-2\tabcolsep-1.3333\arrayrulewidth}
          p{\dimexpr.5\linewidth-2\tabcolsep-1.3333\arrayrulewidth}
      }
            \vspace{-2pt}\caption{Optimal config.} \label{fig:optimal}
          & \vspace{-2pt}\caption{Inference breakdown.} \label{fig:eval_disagg_brk}
      \end{tabularx}
      \vspace{1pt}
  \end{subfigure}
  \vspace{-12pt}
	\caption{Inference performance analysis.} \label{fig:inf_perf}
  \vspace{5pt}
\end{figure*}

\noindent \textbf{Benchmarks and workloads.}
We selected six benchmarks, encompassing 13 unique workloads, to evaluate performance:
i) DLRM embedding operations (`embed') \cite{naumov2019deep}, ii) relational database (`mariadb') \cite{mariadb},
iii) key-value store (`rocksdb') \cite{rocksdb}, iv) text mining (`pattern') \cite{gnugrep, gnucoreutils},
v) web server (`nginx') \cite{nginx}, and vi) file server (`vsftpd') \cite{vsftpd}.
`embed' performs embedding table lookups and aggregates sparse features. Two variants, `rm1' and `rm2', differ in feature lengths and table entry sizes. `mariadb' uses TPC-H to simulate business database workloads. `rocksdb' performs Get/Put queries on over 100K keys. `pattern' includes `find', `line', and `word' variants, which search text and count lines or words in more than 20K documents. `nginx' has two workloads: `web0/1' for static webpages and `filedown' for video streaming. Finally, `vsftpd' uploads thousands of image files in its `fileup' workload. %Each workload was executed 10 times for reliable results.
Each workload was executed 10 times. Table \ref{tbl:eval_workload} provides details of the tested workloads, including the number of I/O requests, data size, system calls, path walk operations, and TCP packets generated during execution.

\vspace{-8pt}
\subsection{Overall Performance Comparisons}
\label{subsec:eval_overall}
Figure \ref{fig:eval_latency} shows the latency of various ISP models, normalized to \texttt{D-VirtFW}. For clarity, the performance is categorized into six: Network operation times (\emph{Network}), Kernel context switches (\emph{Kernel-ctx}), LBA set handshaking (\emph{LBA-set}), SSD access times (\emph{Storage}), System call and OS stack latency (\emph{System}), and ISP kernel latency (\emph{Compute}).

\noindent \textbf{Programmable-ISP.}
While \texttt{P.ISP-R/V} eliminates data transfer overhead for processing, their overall performance is worse than \texttt{Host}, primarily due to \emph{Kernel-ctx} and \emph{LBA-set} overheads. However, \texttt{P.ISP-R/V} outperforms \texttt{Host} in rocksdb-read and nginx-filedown scenarios. These workloads involve frequent Get operations, where system and data movement significantly impact latency. Since \texttt{P.ISP-R/V} bypass OS and system call overhead by running ISP kernels on the device's bare-metal system, they achieve better performance. Note that \texttt{P.ISP-V} exhibits 13.7\% lower latency than \texttt{P.ISP-R} by avoiding RPC and eliminating the need for network responses. 

\noindent \textbf{ISP-container with OS.}
While \texttt{D-Naive}/\texttt{D-FullOS} removes communication overhead (\emph{Kernel-ctx} + \emph{LBA-set}), they face challenges with computational inefficiencies and data movement due to a full software stack; \texttt{D-FullOS} incurs a 9.3\% higher latency than \texttt{P.ISP-V} due to the overhead handling system calls and the OS with limited computing resources. \texttt{D-Naive} exhibits a 12.8\% slowdown compared to \texttt{D-FullOS}, requiring frequent data transfers between the ISP-container processor and controller complexes.

\noindent \textbf{FW-level containerization (DockerSSD).}
\texttt{D-VirtFW} combines the advantages of full-fledged application execution while avoiding the software stack overhead of \texttt{D-FullOS} and the hardware inefficiencies of \texttt{D-Naive}. It delivers significant performance improvements, outperforming \texttt{P.ISP-R/V}, \texttt{D-Naive}, and \texttt{D-FullOS} by 1.6$\times$, 1.8$\times$, and 1.6$\times$, respectively. By accessing backend flash through $\lambda$FS, \texttt{D-VirtFW} eliminates the need for \emph{LBA-set}, achieving an 8.4\% latency reduction compared to \texttt{P.ISP-R/V}. Moreover, execution parameters are pre-packaged in \texttt{rootfs}, removing \emph{Kernel-ctx} overhead and improving ISP performance by 30.9\% over \texttt{P.ISP-R/V}.

\vspace{-8pt}
\subsection{Disaggregated Computing Storage for LLM}
In this extended evaluation, we investigate whether a computing-enabled storage pool powered by DockerSSDs can effectively serve distributed LLM inference. To test this, we select eight LLMs with diverse architectures and varying model sizes: lamda-137B \cite{azizi2024lamda}, gpt3-175B \cite{floridi2020gpt}, jurassic-178B \cite{lieber2021jurassic}, pangu-200B \cite{ren2023pangu}, gopher-280B \cite{rae2021scaling}, turing-530B \cite{sejnowski2023large}, palm-540B \cite{chowdhery2023palm}, and megatron-1T \cite{narayanan2021efficient}. These models, which typically require distributed systems with tens to hundreds of devices, are evaluated using storage pools composed of 16 to 128 DockerSSDs.

\begin{figure*}[]
  \centering
  \begin{minipage}[]{.59\linewidth}
    \centering
    \begin{subfigure}[]{.33\linewidth}
      \includegraphics[width=1\linewidth]{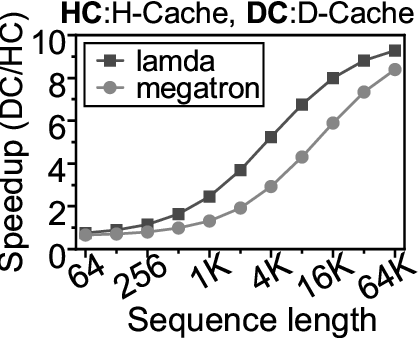}
      \vspace{-8pt} \caption{Model size.} \label{fig:eval_disagg_ss_ms}
    \end{subfigure}
    \begin{subfigure}[]{.65\linewidth}
      \includegraphics[width=1\linewidth]{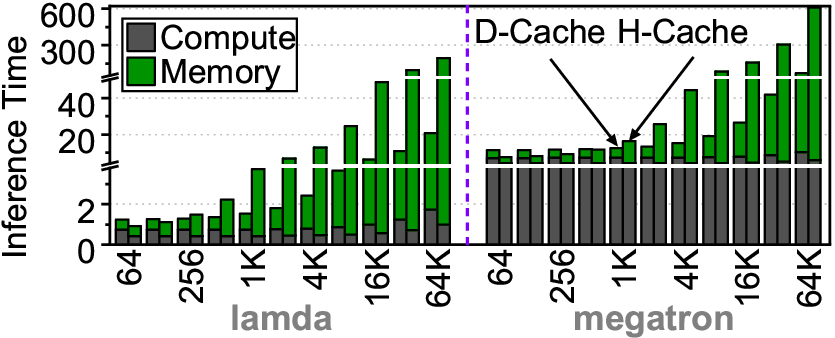}
      \vspace{-8pt} \caption{Breakdown (sequence length).} \label{fig:eval_disagg_ss_sl}
    \end{subfigure}
  \end{minipage}
  \begin{minipage}[]{.39\linewidth}
    \centering
    \begin{subfigure}[]{.49\linewidth}
      \includegraphics[width=1\linewidth]{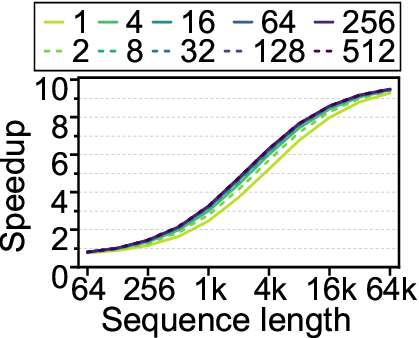}
      \vspace{-7pt} \caption{lamda (137B).} \label{fig:eval_lamda}
    \end{subfigure}
    \begin{subfigure}[]{.49\linewidth}
      \includegraphics[width=1\linewidth]{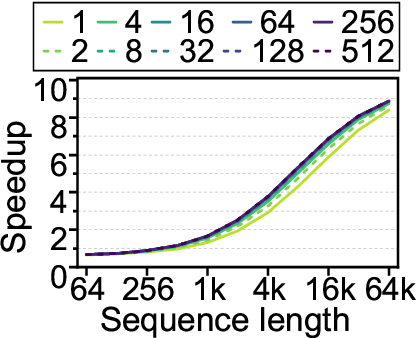}
      \vspace{-7pt} \caption{megatron (1T).} \label{fig:eval_megatron}
    \end{subfigure}
  \end{minipage}
  \vspace{18pt}
  \caption{Sensitivity test.} \label{fig:eval_sensitivity}
\end{figure*}

\noindent \textbf{Resource disaggregation models.}
We prepared three disaggregated configurations based on the location serving distributed inferences (Host or DockerSSD) and the presence of a KV cache (NoCache or Cache). In \texttt{H-NoCache}, distributed inferences are performed across multiple hosts (matching the number of DockerSSDs, 16$\sim$128), each with 64GB of local DRAM. 
This configuration lacks a KV cache due to insufficient DRAM capacity to store KV vectors, but all data resides in local DRAM. In \texttt{H-Cache}, each host uses external storage (400GB SSD) combined with DRAM via Linux swap to support the KV cache. In addition to caching KV data, all other data is also maintained in memory. \texttt{D-Cache} employs DockerSSDs instead of hosts, with each DockerSSD (400GB storage capacity).

\noindent \textbf{Methodology and benchmarks.}
We used an open-source simulator to evaluate distributed inference performance for LLMs by varying parallelization techniques, optimization methods, and system configurations \cite{isaev2023calculon}. Since the existing simulator only supports basic LLM structures and lacks KV cache functionality, we developed an analytical model for KV cache and integrated it into the simulator. We also enhanced it to evaluate performance under different degrees of parallelism (data, tensor, and pipeline) based on GPU counts and batch sizes, identifying the optimal configuration by selecting the scenario with the shortest execution time.

Figure \ref{fig:optimal} presents the optimal parallelism values for each disaggregation model with a sequence length of 32K and a batch size of 1 per GPU. In \texttt{H-NoCache} and \texttt{D-NoCache}, where per-layer computation is heavy, pipeline parallelism minimizes inference time. In contrast, \texttt{H-Cache} and \texttt{D-Cache} leverage KV cache to reduce per-layer computation, making tensor parallelism the most efficient for minimizing total inference time.

\noindent \textbf{Performance analysis.}
Figure \ref{fig:eval_disagg_brk} breaks down distributed LLM inference performance into two components: \texttt{Compute}, which measures the time spent on core operations like matrix and vector multiplications, and \texttt{Memory}, which includes the time required to read input data and write output data during inference. The comparison between \texttt{H-NoCache} and \texttt{D-NoCache} highlights the performance difference based on where distributed inference is performed. DockerSSD achieves distributed inference with only a 1.7$\times$ performance degradation compared to the host. This gap is primarily due to the computational capability difference between the host and DockerSSD.

The differences between \texttt{H-NoCache/H-Cache} and \texttt{D-NoCache/D-Cache}, demonstrate the significant benefits of KV cache techniques. Expanding memory capacity to enable KV caching greatly boosts performance. Specifically, \texttt{H-Cache} achieves a 421$\times$ performance gain over \texttt{H-NoCache}, while \texttt{D-Cache} achieves a remarkable 4.6K$\times$ improvement over \texttt{D-NoCache}. The performance improvement with DockerSSD comes from its ability to access flash memory as local memory, unlike the host, which relies on software-based memory extensions like swap. This advantage makes DockerSSD particularly well-suited for KV cache-based methods, which replace computation with memory storage. As a result, \texttt{D-Cache} outperforms \texttt{H-Cache} by 7.9$\times$ and \texttt{H-NoCache} by 3.2K$\times$ in distributed inference.

\noindent \textbf{Analysis with varying sequence lengths.}
The KV cache size varies significantly with sequence length. Without a KV cache, inference for a sequence length $n$ requires $O(n^2)$ operations. Using an $O(n^2)$-sized KV cache allows reuse of previously computed KV vectors, reducing computational complexity to $O(n)$. This reduction becomes more pronounced as sequence length increases, emphasizing DockerSSD's advantage in efficiently accommodating large KV caches for improved LLM distributed inference. We evaluated DockerSSD's benefits across varying sequence lengths using lamda (smallest model) and megatron (largest model). Figure \ref{fig:eval_disagg_ss_ms} shows that for longer sequences, \texttt{D-Cache} achieves faster inference than \texttt{H-Cache}. For shorter sequences (Figure \ref{fig:eval_disagg_ss_sl}), computation time dominates, and DockerSSD's slower processing speed (2.2GHz vs. 3.8GHz) results in roughly 60\% of host performance. As sequences grow, memory time becomes more significant. Beyond a certain point, DockerSSD's reduced memory overhead outweighs its computational disadvantage. This crossover occurs at a sequence length of 256 for lamda and 1,024 for megatron, where DockerSSD begins outperforming the host.
With longer sequences, the speedup converges to around 9.5$\times$, reflecting the maximum benefit from eliminating the swap overhead in \texttt{H-Cache}. Smaller models exhibit greater speedup for the same sequence length, as larger models allocate more time to MLPs, reducing the impact of KV cache improvements on attention layers. 

\noindent \textbf{Sensitivity test for batch sizes.}
The KV cache size is also influenced by batch size. Figure \ref{fig:eval_lamda}, \ref{fig:eval_megatron} presents the sensitivity analysis for varying batch sizes, using lamda and megatron as representatives, as in the previous evaluation. While larger batch sizes can enhance processing speed leveraging GPU parallelism, the increased memory bandwidth demand from KV caching may create bottlenecks, potentially reducing overall performance. Consequently, smaller batch sizes are often preferable for longer sequence lengths.
As the batch size increases from 1 to 512 for the same sequence length, KV cache memory usage grows linearly with the batch size. However, \texttt{D-Cache} shows only modest improvement over \texttt{H-Cache}, with a maximum speedup of 1.3$\times$ for lamda and megatron.

%-------------------------------------------------------------------------------

%-------------------------------------------------------------------------------
\section{CONCLUSION}
\label{sec:conclusion}
We introduced \emph{DockerSSD}, a novel ISP model that integrates containers and lightweight firmware to enable efficient, containerized data processing directly within SSDs. Leveraging \emph{Ethernet over NVMe} and \emph{Virtual Firmware}, DockerSSD addresses challenges in ISP adaptation and disaggregation, enabling high-performance, computing-enabled storage pools. Our evaluations demonstrate significant reductions in host overhead and enhanced distributed inference performance for large-scale services. These results position DockerSSD as a transformative solution for scalable storage systems in disaggregated infrastructures.
\vspace{-92pt}
%-------------------------------------------------------------------------------

%-------------------------------------------------------------------------------
\vspace*{70pt}
\def\refname{REFERENCES}
\renewcommand\refname{\section{REFERENCES}}
\balance
\bibliographystyle{unsrt}
\bibliography{reference}

\begin{thebibliography}{10}

\bibitem{kang2013enabling}
Yangwook Kang, Yang-suk Kee, Ethan~L Miller, and Chanik Park.
\newblock {Enabling cost-effective data processing with smart SSD}.
\newblock In {\em 2013 IEEE 29th symposium on mass storage systems and technologies (MSST)}, pages 1--12. IEEE, 2013.

\bibitem{lee2014accelerating}
Young-Sik Lee, Luis~Cavazos Quero, Youngjae Lee, Jin-Soo Kim, and Seungryoul Maeng.
\newblock Accelerating external sorting via $\{$On-the-fly$\}$ data merge in active $\{$SSDs$\}$.
\newblock In {\em 6th USENIX Workshop on Hot Topics in Storage and File Systems (HotStorage 14)}, 2014.

\bibitem{seshadri2014willow}
Sudharsan Seshadri, Mark Gahagan, Sundaram Bhaskaran, Trevor Bunker, Arup De, Yanqin Jin, Yang Liu, and Steven Swanson.
\newblock {Willow: A User-Programmable SSD}, OSDI'14.

\bibitem{gu2016biscuit}
Boncheol Gu, Andre~S. Yoon, Duck-Ho Bae, Insoon Jo, Jinyoung Lee, Jonghyun Yoon, Jeong-Uk Kang, Moonsang Kwon, Chanho Yoon, Sangyeun Cho, Jaeheon Jeong, and Duckhyun Chang.
\newblock {Biscuit: A framework for Near-data Processing of Big Data Workloads}, ISCA'16.

\bibitem{kim2016storage}
Sungchan Kim, Hyunok Oh, Chanik Park, Sangyeun Cho, Sang-Won Lee, and Bongki Moon.
\newblock {In-storage processing of database scans and joins}.
\newblock {\em Information Sciences}, 327:183--200, 2016.

\bibitem{jo2016yoursql}
Insoon Jo, Duck-Ho Bae, Andre~S Yoon, Jeong-Uk Kang, Sangyeun Cho, Daniel~DG Lee, and Jaeheon Jeong.
\newblock Yoursql: a high-performance database system leveraging in-storage computing.
\newblock {\em Proceedings of the VLDB Endowment}, pages 924--935, 2016.

\bibitem{jin2017kaml}
Yanqin Jin, Hung-Wei Tseng, Yannis Papakonstantinou, and Steven Swanson.
\newblock Kaml: A flexible, high-performance key-value ssd.
\newblock In {\em 2017 IEEE International Symposium on High Performance Computer Architecture (HPCA)}, pages 373--384. IEEE, 2017.

\bibitem{koo2017summarizer}
Gunjae Koo, Kiran~Kumar Matam, Te~I, HV~Krishna~Giri Narra, Jing Li, Hung-Wei Tseng, Steven Swanson, and Murali Annavaram.
\newblock Summarizer: trading communication with computing near storage.
\newblock In {\em Proceedings of the 50th Annual IEEE/ACM International Symposium on Microarchitecture}, 2017.

\bibitem{jun2018grafboost}
Chun-Yi Liu, Yunju Lee, Wonil Choi, and Myoungsoo Jung.
\newblock Prolonging 3d nand ssd lifetime via read latency relaxation.
\newblock In {\em In Proceedings of the 26th ACM International Conference on Architectural Support for Programming Languages and Operating Systems}, pages 730--742. IEEE, 2021.

\bibitem{torabzadehkashi2019catalina}
Mahdi Torabzadehkashi, Siavash Rezaei, Ali Heydarigorji, Hosein Bobarshad, Vladimir Alves, and Nader Bagherzadeh.
\newblock Catalina: In-storage processing acceleration for scalable big data analytics.
\newblock In {\em 2019 27th Euromicro International Conference on Parallel, Distributed and Network-Based Processing (PDP)}, pages 430--437. IEEE, 2019.

\bibitem{ruan2019insider}
Zhenyuan Ruan, Tong He, and Jason Cong.
\newblock $\{$INSIDER$\}$: Designing $\{$In-Storage$\}$ computing system for emerging $\{$High-Performance$\}$ drive.
\newblock In {\em 2019 USENIX Annual Technical Conference (USENIX ATC 19)}, pages 379--394, 2019.

\bibitem{mansouri2022genstore}
Nika Mansouri~Ghiasi, Jisung Park, Harun Mustafa, Jeremie Kim, Ataberk Olgun, Arvid Gollwitzer, Damla Senol~Cali, Can Firtina, Haiyu Mao, Nour Almadhoun~Alserr, et~al.
\newblock Genstore: A high-performance in-storage processing system for genome sequence analysis.
\newblock In {\em Proceedings of the 27th ACM International Conference on Architectural Support for Programming Languages and Operating Systems}, pages 635--654, 2022.

\bibitem{gouk2023containerized}
Donghyun Gouk, Miryeong Kwon, Hanyeoreum Bae, and Myoungsoo Jung.
\newblock Containerized in-storage processing model and hardware acceleration for fully-flexible computational ssds.
\newblock {\em IEEE Computer Architecture Letters}, 2023.

\bibitem{gouk2024dockerssd}
Donghyun Gouk, Miryeong Kwon, Hanyeoreum Bae, and Myoungsoo Jung.
\newblock Dockerssd: Containerized in-storage processing and hardware acceleration for computational ssds.
\newblock In {\em 2024 IEEE International Symposium on High-Performance Computer Architecture (HPCA)}, pages 379--394. IEEE, 2024.

\bibitem{kvssd}
Samsung.
\newblock {Samsung Key Value SSD enables High Performance Scaling}.
\newblock \url{https://www.samsung.com/semiconductor/global.semi.static/Samsung_Key_Value_SSD_enables_High_Performance_Scaling-0.pdf}, 2017.

\bibitem{im2020pink}
Junsu Im, Jinwook Bae, Chanwoo Chung, Sungjin Lee, et~al.
\newblock $\{$PinK$\}$: High-speed in-storage key-value store with bounded tails.
\newblock In {\em 2020 USENIX Annual Technical Conference (USENIX ATC 20)}, pages 173--187, 2020.

\bibitem{issd}
Argonboards.
\newblock {LS2088A Intelligent-SSD Card}.
\newblock \url{https://www.argonboards.com/ls2088a-intelligent-ssd-card}, 2022.

\bibitem{kwon2022vigil}
Miryeong Kwon, Seungjun Lee, Hyunkyu Choi, Jooyoung Hwang, and Myoungsoo Jung.
\newblock $\{$Vigil-KV$\}$:$\{$Hardware-Software$\}$$\{$Co-Design$\}$ to integrate strong latency determinism into $\{$Log-Structured$\}$ merge $\{$Key-Value$\}$ stores.
\newblock In {\em 2022 USENIX Annual Technical Conference (USENIX ATC 22)}, pages 755--772, 2022.

\bibitem{duffy2023dotori}
Carl Duffy, Jaehoon Shim, Sang-Hoon Kim, and Jin-Soo Kim.
\newblock Dotori: A key-value ssd based kv store.
\newblock {\em Proceedings of the VLDB Endowment}, 16(6):1560--1572, 2023.

\bibitem{park2023kv}
Inhyuk Park, Qing Zheng, Dominic Manno, Soonyeal Yang, Jason Lee, David Bonnie, Bradley Settlemyer, Youngjae Kim, Woosuk Chung, and Gary Grider.
\newblock Kv-csd: A hardware-accelerated key-value store for data-intensive applications.
\newblock In {\em 2023 IEEE International Conference on Cluster Computing (CLUSTER)}, pages 132--144, 2023.

\bibitem{azizi2024lamda}
Seyedarmin Azizi, Souvik Kundu, and Massoud Pedram.
\newblock Lamda: Large model fine-tuning via spectrally decomposed low-dimensional adaptation.
\newblock {\em arXiv preprint arXiv:2406.12832}, 2024.

\bibitem{floridi2020gpt}
Luciano Floridi and Massimo Chiriatti.
\newblock Gpt-3: Its nature, scope, limits, and consequences.
\newblock {\em Minds and Machines}, 30:681--694, 2020.

\bibitem{lieber2021jurassic}
Opher Lieber, Or~Sharir, Barak Lenz, and Yoav Shoham.
\newblock Jurassic-1: Technical details and evaluation.
\newblock {\em White Paper. AI21 Labs}, 1(9):1--17, 2021.

\bibitem{ren2023pangu}
Xiaozhe Ren, Pingyi Zhou, Xinfan Meng, Xinjing Huang, Yadao Wang, Weichao Wang, Pengfei Li, Xiaoda Zhang, Alexander Podolskiy, Grigory Arshinov, et~al.
\newblock Pangu-$\{$$\backslash$Sigma$\}$: Towards trillion parameter language model with sparse heterogeneous computing.
\newblock {\em arXiv preprint arXiv:2303.10845}, 2023.

\bibitem{rae2021scaling}
Jack~W Rae, Sebastian Borgeaud, Trevor Cai, Katie Millican, Jordan Hoffmann, Francis Song, John Aslanides, Sarah Henderson, Roman Ring, Susannah Young, et~al.
\newblock Scaling language models: Methods, analysis \& insights from training gopher.
\newblock {\em arXiv preprint arXiv:2112.11446}, 2021.

\bibitem{sejnowski2023large}
Terrence~J Sejnowski.
\newblock Large language models and the reverse turing test.
\newblock {\em Neural computation}, 35(3):309--342, 2023.

\bibitem{chowdhery2023palm}
Aakanksha Chowdhery, Sharan Narang, Jacob Devlin, Maarten Bosma, Gaurav Mishra, Adam Roberts, Paul Barham, Hyung~Won Chung, Charles Sutton, Sebastian Gehrmann, et~al.
\newblock Palm: Scaling language modeling with pathways.
\newblock {\em Journal of Machine Learning Research}, 24(240):1--113, 2023.

\bibitem{narayanan2021efficient}
Deepak Narayanan, Mohammad Shoeybi, Jared Casper, Patrick LeGresley, Mostofa Patwary, Vijay Korthikanti, Dmitri Vainbrand, Prethvi Kashinkunti, Julie Bernauer, Bryan Catanzaro, et~al.
\newblock Efficient large-scale language model training on gpu clusters using megatron-lm.
\newblock In {\em Proceedings of the international conference for high performance computing, networking, storage and analysis}, pages 1--15, 2021.

\bibitem{jung2020openexpress}
Myoungsoo Jung.
\newblock {OpenExpress: Fully Hardware Automated Open Research Framework for Future Fast NVMe Devices}, USENIX ATC'20.

\bibitem{shadley2018deployment}
Scott Shadley.
\newblock Deployment of in-storage compute.
\newblock In {\em Storage Developer Conference}, 2018.

\bibitem{jung2013revisiting}
Myoungsoo Jung and Mahmut Kandemir.
\newblock Revisiting widely held ssd expectations and rethinking system-level implications.
\newblock {\em ACM SIGMETRICS Performance Evaluation Review}, 41(1):203--216, 2013.

\bibitem{matam2019graphssd}
Kiran~Kumar Matam, Gunjae Koo, Haipeng Zha, Hung-Wei Tseng, and Murali Annavaram.
\newblock Graphssd: graph semantics aware ssd.
\newblock In {\em Proceedings of the 46th international symposium on computer architecture}, pages 116--128, 2019.

\bibitem{kim2023decoupled}
Jiho Kim, Myoungsoo Jung, and John Kim.
\newblock Decoupled ssd: Rethinking ssd architecture through network-based flash controllers.
\newblock In {\em Proceedings of the 50th Annual International Symposium on Computer Architecture}, pages 1--13, 2023.

\bibitem{sun2023leaftl}
Jinghan Sun, Shaobo Li, Yunxin Sun, Chao Sun, Dejan Vucinic, and Jian Huang.
\newblock Leaftl: A learning-based flash translation layer for solid-state drives.
\newblock In {\em Proceedings of the 28th ACM International Conference on Architectural Support for Programming Languages and Operating Systems, Volume 2}, pages 442--456, 2023.

\bibitem{layerscapeprocessors}
NXP.
\newblock {Layerscape 2088A and 2048A Multicore Communications Processors}.
\newblock \url{https://www.nxp.com/products/LS2088A}.

\bibitem{microsemipm8609}
Microchip~Technology Inc.
\newblock {Microsemi PM8609 NVMe2032 Controller}.
\newblock \url{https://www.microchip.com/en-us/product/pm8609}.

\bibitem{nvmexpress}
{NVM Express, Inc}.
\newblock {NVM Express}.
\newblock \url{https://nvmexpress.org/wp-content/uploads/NVM-Express-Base-Specification-Revision-2.2-2025.03.11-Ratified.pdf}.

\bibitem{linuxcgroups}
{Linux Man}.
\newblock {Linux Control Groups v1}.
\newblock \url{https://www.kernel.org/doc/Documentation/cgroup-v1/cgroups.txt}.

\bibitem{linuxnamespaces}
{Linux Man}.
\newblock {Linux Namespaces}.
\newblock \url{https://man7.org/linux/man-pages/man7/namespaces.7.html}.

\bibitem{merkel2014docker}
Dirk Merkel.
\newblock {Docker: Lightweight Linux Containers for Consistent Development and Deployment}.
\newblock {\em Linux journal}, 2014(239), 2014.

\bibitem{docker}
{Docker, Inc}.
\newblock {docker}.
\newblock \url{https://www.docker.com/}.

\bibitem{dockercompose}
{Docker, Inc}.
\newblock {Docker Compose}.
\newblock \url{https://docs.docker.com/compose/}.

\bibitem{kubernetes}
{Cloud Native Computing Foundation}.
\newblock {Kubernetes}.
\newblock \url{https://kubernetes.io/}.

\bibitem{virtexup}
Xilinx.
\newblock {Xilinx Virtex UltraScale+}.
\newblock \url{https://www.xilinx.com/products/silicon-devices/fpga/virtex-ultrascale-plus.html}.

\bibitem{jung2017simplessd}
Myoungsoo Jung, Jie Zhang, Ahmed Abulila, Miryeong Kwon, Narges Shahidi, John Shalf, Nam~Sung Kim, and Mahmut Kandemir.
\newblock {SimpleSSD: Modeling Solid State Drives for Holistic System Simulation}.
\newblock {\em IEEE Computer Architecture Letters}, 17(1), 2017.

\bibitem{binkert2011gem5}
Nathan Binkert, Bradford Beckmann, Gabriel Black, Steven~K. Reinhardt, Ali Saidi, Arkaprava Basu, Joel Hestness, Derek~R. Hower, Tushar Krishna, Somayeh Sardashti, Rathijit Sen, Korey Sewell, Muhammad Shoaib, Nilay Vaish, Mark~D. Hill, and David~A. Wood.
\newblock The gem5 simulator.
\newblock {\em ACM SIGARCH Computer Architecture News}, 39(2), 2011.

\bibitem{naumov2019deep}
Maxim Naumov, Dheevatsa Mudigere, Hao-Jun~Michael Shi, Jianyu Huang, Narayanan Sundaraman, Jongsoo Park, Xiaodong Wang, Udit Gupta, Carole-Jean Wu, Alisson~G Azzolini, et~al.
\newblock Deep learning recommendation model for personalization and recommendation systems.
\newblock {\em arXiv preprint arXiv:1906.00091}, 2019.

\bibitem{mariadb}
{MariaDB Foundation}.
\newblock {MariaDB}.
\newblock \url{https://mariadb.org}.

\bibitem{rocksdb}
{Meta Platforms, Inc.}
\newblock {RocksDB}.
\newblock \url{https://rocksdb.org}.

\bibitem{gnugrep}
{Free Software Foundation, Inc}.
\newblock {GNU grep}.
\newblock \url{https://www.gnu.org/software/grep/.}

\bibitem{gnucoreutils}
{Free Software Foundation, Inc}.
\newblock {GNU coreutils}.
\newblock \url{https://www.gnu.org/software/coreutils/.}

\bibitem{nginx}
{Nginx, Inc}.
\newblock {Nginx}.
\newblock \url{https://www.nginx.com/.}

\bibitem{vsftpd}
{Chris Evans}.
\newblock {vsftpd}.
\newblock \url{https://security.appspot.com/vsftpd.html.}

\bibitem{isaev2023calculon}
Mikhail Isaev, Nic McDonald, Larry Dennison, and Richard Vuduc.
\newblock Calculon: a methodology and tool for high-level co-design of systems and large language models, SC'23.

\end{thebibliography}

\end{document}